\documentclass[11pt]{article}
\pdfoutput=1

\usepackage{euscript}
\usepackage{amssymb}
\usepackage{amsfonts}
\usepackage{amsbsy}
\usepackage{epsfig}
\usepackage{amsthm}
\usepackage{amscd}
\usepackage{amstext}
\usepackage{verbatim}
\usepackage{amsmath}
\usepackage{cancel}
\usepackage{capt-of}
\usepackage{empheq}
\usepackage{subfigure}

\usepackage[pdftex]{hyperref}

\def\ben{\begin{equation}}
\def\een{\end{equation}}
\def\half{{\textstyle{1\over2}}}

   \let\d=\delta 
   
\let\l=\lambda

\def\be{\begin{equation}}
\def\ee{\end{equation}}
\def\beq{\begin{equation}}
\def\eeq{\end{equation}}
\def\ba{\begin{array}}
\def\ea{\end{array}}

\def\dalemb#1#2{{\vbox{\hrule height .#2pt
       \hbox{\vrule width.#2pt height#1pt \kern#1pt
               \vrule width.#2pt}
       \hrule height.#2pt}}}

\newcommand{\bea}{\begin{eqnarray}}
\newcommand{\eea}{\end{eqnarray}}
\newcommand{\tr}{{\rm tr} }

\renewcommand{\eqref}[1]{(\ref{eq:#1})}

\begin{document}

\begin{center}

{ \Large {\bf
Entanglement entropy in two dimensional string theory
}}

\vspace{1cm}

Sean A. Hartnoll and Edward Mazenc

\vspace{1cm}

{\small
{\it Department of Physics, Stanford University, \\
Stanford, CA 94305-4060, USA }}

\vspace{1.6cm}

\end{center}

\begin{abstract}

To understand an emergent spacetime is to understand the emergence of locality.
Entanglement entropy is a powerful diagnostic of locality, because locality leads to a
large amount of short distance entanglement. Two dimensional string theory is among the very simplest
instances of an emergent spatial dimension. We compute the entanglement entropy
in the large $N$ matrix quantum mechanics dual to two dimensional string theory, in the
semiclassical limit of weak string coupling. We isolate a logarithmically large, but finite,
contribution that corresponds to the short distance entanglement of the tachyon field
in the emergent spacetime. From the spacetime point of view, the entanglement is regulated
by a nonperturbative `graininess' of space.

\end{abstract}

\pagebreak
\setcounter{page}{1}

\section{Introduction}

Locality is a key ingredient of conventional quantum field theories. To show that a given theory of quantum gravity exhibits an emergent semiclassical spacetime, an essential aspect to be understood is the emergence of local dynamics in the emergent spacetime. A robust and universal probe of locality is entanglement entropy. In a quantum system with degrees of freedom at all scales, such as a quantum field theory, local interactions imply a large amount of short distance entanglement in the ground state \cite{Bombelli:1986rw, Srednicki:1993im}.
This suggests the following strategy: Suppose we are given a quantum state that is a candidate to describe an emergent spacetime (say, the ground state of ${\mathcal N} = 4$ SYM theory \cite{Maldacena:1997re} or the ground state of BFSS matrix quantum mechanics \cite{Banks:1996vh}). To `find' the locality of the emergent spacetime within this state, that is, to identify the degrees of freedom that interact locally in the emergent spacetime, what we must do is identify strongly entangled degrees of freedom in the state.

In this paper we will implement the above strategy in the simplest possible example of emergent spacetime. This is the emergence of spacetime in two dimensional string theory from the matrix quantum mechanics of a single large $N$ matrix. While relatively simple, this duality can be thought of as the baby cousin of the AdS/CFT correspondence. We will see that it illustrates how quantum gravity can provide nonperturbative cutoffs on UV divergences that appear due to short distance entanglement in quantum field theory.

The low energy (compared to the string scale) effective target space action for two dimensional bosonic string theory takes the form \cite{Callan:1985ia, Natsuume:1994sp}
\be\label{eq:string}
S = \int dt\,dx \, \sqrt{-g} e^{- 2 \Phi} \left(R + 4 \left(\nabla \Phi \right)^2 + 16 - \left(\nabla T \right)^2 + 4 T^2 - 2 \widetilde V(T) \right) \,. 
\ee
The fields are the metric $g$, `tachyon' $T$ and dilaton $\Phi$. This theory has only one propagating degree of freedom, which we can consider to be the tachyon. The background of interest is the linear dilaton solution together with a tachyon condensate
\be\label{eq:back}
g_{\mu\nu} = \eta_{\mu\nu} \,, \qquad \Phi = 2 x + \cdots \,, \qquad T = \mu \, \left(x + \frac{\log \mu}{2} \right) e^{2x} + \cdots \,.
\ee
The dots indicate that the solution can only be trusted in the regime of weak string coupling
\be\label{eq:gs}
g_\text{s} \equiv e^\Phi \approx e^{2 x} \,,
\ee
which becomes small as $x \to - \infty$. The relative coefficients of the two terms in the tachyon profile are fixed by nonlinearities in the tachyon action
\cite{Natsuume:1994sp}. The single free parameter $\mu$ in the background determines the string coupling at the `tachyon wall' ($x \approx - \half \log \mu$) to be $g_\text{eff.} \approx \,\mu^{-1}$. Therefore, $\mu \to \infty$ is a weakly coupled limit in which scattering off the tachyon wall can be described perturbatively.

Our objective in this paper is the following. We will evaluate the entanglement entropy of a spatial interval of length $\Delta x = x_2 - x_1$ that is to the left of the tachyon wall ($2 x \lesssim - \log \mu$) and in the weakly interacting limit $\mu \to \infty$. In this regime the spacetime is semiclassical and so the entanglement entropy of a spatial region makes sense. About the background (\ref{eq:back}) the tachyon is a massless field. The nonlinearities of the tachyon action can still be important close to the tachyon wall, however, and couple linearized fluctuations of the tachyon to the non-translation-invariant tachyon background (\ref{eq:back}). These effects become small in the limit $x \to - \infty$. In this limit, with weak string interactions and weak nonlinearities, we expect to find the entanglement entropy of a two dimensional massless scalar \cite{Holzhey:1994we}
\be\label{eq:13}
S = \frac{1}{3} \log \frac{\Delta x}{\epsilon} \,, \qquad (x \to - \infty, \; \Delta x \text{ fixed}) \,.
\ee
Here $\epsilon$ is a UV cutoff. The semiclassical spacetime computation of this quantity is simply divergent and cannot see the cutoff. We will instead obtain the result (\ref{eq:13}), complete with an explicit cutoff, from the full underlying quantum state out of which the spacetime emerges.

Our result is given in equation (\ref{eq:final}) below. This result includes the effects of the broken translation invariance. To compare with (\ref{eq:13}) we can again take the limit $x \to - \infty$.
Our result simplifies to
\be\label{eq:result}
S = \frac{1}{3} \log \frac{\mu \, \Delta x}{\sqrt{g_\text{s}(x_1) \, g_\text{s}(x_2)}} \,, \qquad (x \to - \infty, \; \Delta x \text{ fixed}) \,.
\ee
A similar result (with $x_1 \approx x_2$) has been given some time ago in the prescient papers \cite{das1,das2}. Our result (\ref{eq:final}) is more precise and we believe our treatment is more transparent, although the essential physics is the same as that discussed in \cite{das1,das2}. The finite answer for the entanglement indicates a fundamental nonperturbative `graininess' of spacetime at the scale set by the string coupling $g_\text{s}= e^{\Phi} \ll 1$. It is natural to think of this scale as the de Broglie wavelength of the D-particles in the theory, whose condensation has created the spacetime \cite{john, Martinec:2004td}. Also, because the dominant short distance entanglement comes from the boundaries of the region $[x_1,x_2]$, it is natural that the cutoffs in (\ref{eq:result}) are set by the string coupling evaluated at the endpoints $x_1$ and $x_2$.

The worldsheet string theory describing the background (\ref{eq:back}) is Liouville theory coupled to $c=1$ matter. By discretizing the worldsheet, this theory is equivalent to a certain matrix quantum mechanics tuned to a critical point in a double scaling limit, as reviewed in \cite{Klebanov:1991qa, Ginsparg:1993is, Polchinski:1994mb, Martinec:2004td}. The matrix quantum mechanics has the action
\be\label{eq:MQM}
S = \beta N \int dt \, \tr \left[\frac{1}{2} \dot M^2 + V(M) \right] \,.
\ee
Here $M$ is an $N \times N$ Hermitian matrix and $\beta$ is a coupling. We can restrict to singlet states, for instance by gauging the time derivative. The singlet states can be described by the eigenvalues of the matrix $M$. These eigenvalues experience the usual Vandermonde repulsion. For the quantum mechanics (\ref{eq:MQM}) of a single matrix, the Vandermonde repulsion is equivalent to Pauli exclusion of fermions. The dynamics of the eigenvalues can therefore be formulated as the dynamics of $N$ non-interacting spinless fermions with second-quantized Hamiltonian
\be\label{eq:F1}
H = \beta N \int d\lambda \left[\frac{1}{2 (\beta N)^2} \frac{d \Psi^\dagger}{d\lambda}\frac{d \Psi}{d\lambda} + V(\lambda) \Psi^\dagger \Psi \right] \,.
\ee
The emergent spacetime in which the string theory (\ref{eq:string}) lives is in fact the Fermi surface of this theory, and the tachyon field is the bosonization of the fermion dynamics, describing density fluctuations of the Fermi surface. However, in order to obtain a continuum limit of the string theory worldsheet, one must take a certain double scaling limit of the fermion theory (\ref{eq:F1}).

The fermions fill up a well in the potential $V(\lambda)$. At some critical $\beta_c$, the Fermi energy $\epsilon_F$ comes close to a local maximum $\epsilon_c$ of the potential. In this regime there is a logarithmic divergence in the density of states. These large numbers of states must be populated to generate the worldsheet continuum limit. The double scaling limit we take is then
 $N \to \infty$, $\beta \to \beta_c$, holding $\mu \equiv \beta N (\epsilon_c - \epsilon_F)$ fixed. In this limit the fermion Hamiltonian describes fermions in an inverted harmonic oscillator potential:
 \be\label{eq:F2}
 H - \beta N \epsilon_F = \int d\lambda \left[\frac{1}{2} \frac{d \Psi^\dagger}{d\lambda}\frac{d \Psi}{d\lambda} - \frac{\lambda^2}{2} \Psi^\dagger \Psi + \mu \Psi^\dagger \Psi \right] \,.
 \ee
We have allowed ourselves to rescale the fermion fields and to rescale the coordinates as we zoom into the region close to the local maximum of the potential (see e.g. \cite{Polchinski:1994mb}). The ground state of (\ref{eq:F2}) is a Fermi sea in the inverted oscillator potential. The Fermi sea is populated up to a distance $\mu$ from the top of the potential. We will work with the original bosonic (and ultimately unstable) version of the theory in which the Fermi sea is only populated on one side (the right hand side) of the local maximum. The results can surely be adapted to the stable fermionic theory in which both sides are populated \cite{Takayanagi:2003sm, Douglas:2003up}.

Extensive work on this duality has mapped various quantities between the fermion description (\ref{eq:F2}) and the spacetime description (\ref{eq:string}). The basic quantities that are computed are the S-matrix elements for (in one frame) the scattering of tachyons off the tachyon wall which is equivalent to (in the other frame) the scattering of deformations of the Fermi surface off the inverted oscillator potential \cite{Klebanov:1991qa, Ginsparg:1993is, Polchinski:1994mb, Martinec:2004td}. We will only need two results from this past work. The first is that the quantity $\mu$ appearing in (\ref{eq:F2}) and in the background (\ref{eq:back}) is the same. In particular, this means that in order to access the weakly interacting semiclassical spacetime regime we must take $\mu \to \infty$. As we will see shortly, this corresponds to taking a WKB limit of the fermionic wavefunctions.

The second result we will take from past work is that, in the weakly coupled regime as $\mu \to \infty$, the fermionic coordinate $\lambda$ in (\ref{eq:F2}) is related to the spacetime coordinate $x$ in (\ref{eq:string}) by the `time of flight' relation \cite{DJ}\footnote{It is known that the relation (\ref{eq:exp}) is not precise \cite{MS}. The nonlocal nature of the full map between $x$ and $\lambda$ is due to the existence of additional massive string states at certain discrete momenta \cite{MS}. It is appropriate to use the local map (\ref{eq:exp}), as we will do, to reproduce the contribution of the spacetime tachyon field to the entanglement entropy of a spatial region in the weakly coupled limit $\mu \to \infty$.}
\be\label{eq:exp}
x = - \frac{1}{\sqrt{2}} \int_{\lambda_\star(\mu)}^\lambda \frac{d \lambda'}{\sqrt{-V(\lambda') - \mu}} = - \log \frac{\lambda + \sqrt{\lambda^2 - 2 \mu}}{\sqrt{2 \mu}}\,.
\ee
In the second relation we have used $V(\lambda) = -\lambda^2/2$ and hence the turning point $\lambda_\star(\mu) = \sqrt{2 \mu}$.
In fact, one can also take our results to be an alternative {\it derivation} of the relation (\ref{eq:exp}), based on the need to recover the short distance entanglement (\ref{eq:13}) of a one dimensional massless scalar field from the entanglement of the matrix eigenvalues. Note that in (\ref{eq:exp}) we are sending positive $\lambda$, to the right of the maximum of the potential, to negative spacetime coordinate $x$.

Given the rather explicit map from the eigenvalue ($=$ fermion) description to the emergent spacetime in this particular duality, there is a natural guess of which matrix quantum mechanics degrees of freedom will correspond to a given spatial region in spacetime. Namely, one expects that the eigenvalues taking some continuous range of values will map onto a region of spacetime according to the relation (\ref{eq:exp}). With this in mind, we will proceed to compute the entanglement entropy of an interval in the theory of non-interacting fermions in an inverted harmonic oscillator potential (\ref{eq:F2}). As discussed above, we expect this to give a complete and manifestly UV finite computation of the entanglement entropy that is not accessible directly from the spacetime perspective of the action (\ref{eq:string}).

\section{Entanglement and density fluctuations}

The entanglement entropy of non-interacting fermions in a region $A$ can be expressed as a sum over cumulants of the particle number distribution \cite{levitov, song}
\be\label{eq:cumu}
S_A = \frac{\pi^2}{3} V_A^{(2)} + \frac{\pi^4}{45} V_A^{(4)} + \frac{2 \pi^6}{945} V_A^{(6)} + \cdots \,.
\ee
Here
\be
V_A^{(m)} = \left. \left( - i \frac{d}{d\lambda} \right)^m \log \left\langle e^{i \lambda N_A} \right\rangle \right|_{\lambda = 0} \,,
\ee
with the integrated density operator in the region $A$ given by
\be
N_A = \int_A d\lambda \, n(\lambda) \,.
\ee
The expression for the entanglement entropy in terms of density fluctuations is ultimately derived from
the explicitly known reduced density matrix of a region for a system of non-interacting fermions. In particular, the reduced density matrix and entanglement entropy of a region can be expressed both in terms of the matrix of fermion two point correlation functions \cite{peschel} and also in terms of the matrix of fermion wavefunction overlaps in the region \cite{kl}. We also found the further discussion in \cite{calab} helpful.

The expansion of the entanglement entropy in terms of the density cumulants in (\ref{eq:cumu}) will be especially useful for us for the following reason. It has been found that the leading singular behavior of the entanglement entropy in the limit of large fermion occupation number is determined by the second cumulant alone \cite{calab1, calab2}. We will see shortly below that the large $\mu$ limit of interest to us is a WKB limit for the fermions and hence indeed corresponds to large fermion occupation number. Therefore, we can
expect that to leading order in the large $\mu$ limit
\be\label{eq:dens}
S_A = \frac{\pi^2}{3} V_A^{(2)} = \frac{\pi^2}{3} \int_A d\l \,d\l' \Big(\left\langle n(\l) n(\l') \right\rangle - \left\langle n(\l) \right\rangle \left\langle n(\l') \right\rangle  \Big) \,.
\ee
Actually, this result -- that one can restrict to $V_A^{(2)}$ to leading order -- has not been shown for noninteracting electrons in an arbitrary potential. In the additional limit in which the size of the entangling region becomes small relative to the scale of variation in the potential (this is the limit considered in \cite{das1,das2}), then the cancellations described in \cite{calab1} for the short distance engagement will occur independently of the form of potential. We will see below that we can do better than this, by noting that in the WKB limit the singular contributions to all of the cumulants $V_A^{(m)}$ are functions of the time of flight variable (\ref{eq:exp}).

The density operator is given by the second quantized fermionic field operators $\Psi$ as
\be
n(t,\l) = \Psi^\dagger(t,\lambda) \Psi(t,\lambda) \,.
\ee
The field operators can be expressed in terms of creation and annihilation operators weighted by energy eigenfunctions of the associated single particle Hamiltonian:
\be\label{eq:psi}
\Psi(t,\lambda) = \int_{-\infty}^\infty d\nu \, e^{i \nu t} a(\nu) \psi_\nu(\l) \,.
\ee
That is
\be
- \frac{1}{2} \frac{d^2 \psi_\nu}{d \l^2} + V(\l) \psi_\nu = - \nu \psi_\nu \,.
\ee
Note that $\nu$ is minus the energy. Similarly $\mu$ will be the chemical potential measured downwards towards negative energies, so that the state satisfies
\be
\begin{array}{cc}
a_\nu | \mu \rangle = 0 \,, & \text{if} \quad \nu < \mu \,, \\
a^\dagger_\nu | \mu \rangle = 0 \,, & \text{if} \quad \nu > \mu \,.
\end{array}
\ee
In (\ref{eq:psi}) one often includes a sum over parities of the wavefunctions (e.g. \cite{Moore:1991sf}).
However, we are interested in a background in which the Fermi sea is filled only on one side of the local maximum.
The relevant states therefore have wavefunctions of the form $\psi = (\psi^+ + \psi^-)/\sqrt{2}$, that vanish on one side of the potential in the WKB limit.
The density correlation in (\ref{eq:dens}) is at equal times. Standard manipulations then give \cite{Moore:1991sf}
\be\label{eq:nnnn}
\left\langle n(\l) n(\l') \right\rangle - \left\langle n(\l) \right\rangle \left\langle n(\l') \right\rangle = 
\int_\mu^\infty d\nu_1 \, \psi_{\nu_1}(\lambda) \psi_{\nu_1}(\lambda') \int_{-\infty}^\mu d\nu_2 \,
\psi_{\nu_2}(\lambda) \psi_{\nu_2}(\lambda') \,.
\ee
We have used the fact that the wavefunctions will be real.

\section{Computation of the particle variation}

It is clear that to evaluate the entanglement entropy, following equations (\ref{eq:dens}) and (\ref{eq:nnnn}) above, we need to compute the integrals
\be\label{eq:quartic}
3 S_A = \pi^2 \int_\mu^\infty d\nu_1 \int_{-\infty}^\mu d\nu_2 \left(\int_{\l_1}^{\l_2} d\lambda \, \psi_{\nu_1}(\lambda) \psi_{\nu_2}(\lambda) \right)^2 \,. 
\ee
We have taken the region $A$ to be the interval $[\l_1,\l_2]$. The $\lambda$ integral is immediately performed by noting that
\be\label{eq:trick}
\int_{\l_1}^{\l_2} d\lambda \, \psi_{\nu_1}(\lambda) \psi_{\nu_2}(\lambda) = \frac{1}{2} \frac{1}{\nu_1 - \nu_2} \left[ \frac{ d \psi_{\nu_1}}{d\lambda} \psi_{\nu_2} - \psi_{\nu_1} \frac{d\psi_{\nu_2}}{d\lambda}\right]^{\l_2}_{\l_1} \,.
\ee
This result follows directly from the Schr\"odinger equation
\be\label{eq:SE}
- \frac{1}{2} \frac{d^2\psi_\nu}{d\lambda^2} + V(\lambda) \psi_\nu = - \nu \, \psi_\nu \,.
\ee
In particular, (\ref{eq:trick}) is exact and does not depend on any WKB limit. It also does not depend on the form of the Schr\"odinger potential.

We specialize now to the potential of interest $V(\lambda) = - \lambda^2/2$. The occupied states have $\nu > \mu$ in the
Schr\"odinger equation (\ref{eq:SE}). Therefore $\nu$ is large in the large $\mu$ limit, and the wavefunctions are correctly captured by a WKB limit. Of course, the WKB description will break down sufficiently close to turning points, but we will see that these regions are not important for the quantities we are after. Performing the standard matching across the turning points, the WKB wavefunctions for $\lambda > \sqrt{2 \nu}$ are, up to corrections that are exponentially small at large $\mu$,
\be\label{eq:WKB}
\psi_\nu(\lambda) = \frac{\sqrt{2}}{\sqrt{\pi p}} \sin \left(\int_{\sqrt{2 \nu}}^\lambda p \, d\lambda - \frac{\pi}{4} \right) \,,
\ee
where
\be
p = \sqrt{\lambda^2 - 2 \nu} \,.
\ee
So that
\be
P_\nu(\lambda) \equiv \int_{\sqrt{2 \nu}}^\lambda p \, d\lambda = \frac{1}{2} \left( \lambda \sqrt{\lambda^2 - 2 \nu} - 2 \nu \log \frac{\lambda + \sqrt{\lambda^2 - 2 \nu}}{\sqrt{2 \nu}}\right) \,.
\ee
The normalization of (\ref{eq:WKB}) follows, for instance, from taking the WKB limit of the full delta function normalized wavefuntions written in terms of confluent hypergeometric functions (e.g. \cite{Moore:1991sf}). Our wavefunctions have an extra $\sqrt{2}$ in their normalization relative to \cite{Moore:1991sf}, as we are restricting to modes that are zero on one side of the local maximum.

The integrand obtained from substituting the WKB wavefunctions (\ref{eq:WKB}) into (\ref{eq:trick}), and then squaring as required by (\ref{eq:quartic}), can be written as exponentials of linear combinations of $P_{\nu_1}(\l_1), P_{\nu_2}(\l_1), P_{\nu_1}(\l_2), P_{\nu_2}(\l_2)$. Many of these terms are oscillatory. Upon integration, the oscillating terms experience significant cancellations. There are no stationery phase points in the region of integration and hence the oscillating terms vanish in the WKB limit $\mu \to \infty$. In order for the oscillations to be tamed, the exponents must cancel. This can happen in two ways. Firstly the exponent can cancel exactly, to leave a non-oscillating term. Secondly, the exponent can take forms such as $P_{\nu_1}(\l_1) - P_{\nu_2}(\l_1)$. These terms oscillate, but for $\nu_1 \sim \nu_2$ the two terms almost cancel and the oscillations are slowed down.

To do the necessary integrals, it is convenient to make the change of variables
\be\label{eq:nudelta}
\nu_1 = \mu + \nu + \delta \,, \qquad \nu_2 = \mu + \nu - \delta \,.
\ee
The integrals become
\be
\int_\mu^\infty d\nu_1 \int_{-\infty}^\mu d\nu_2 = 2 \int_0^\infty d\delta \int_{-\delta}^\delta d \nu \,.
\ee
In these variables we see that when $\nu_1 \sim \nu_2$, then $\delta$ is small. Furthermore, we see that when $\delta$ is small then $\nu$ is also necessarily small. Therefore, to firstly try to isolate the oscillating contribution with $\nu_1 \sim \nu_2$, we can expand the integrand in (\ref{eq:quartic}) for $\nu_1$ and $\nu_2$ close to $\mu$. Clearly, this is a contribution from fermions close to the Fermi surface. We find that (\ref{eq:quartic}) becomes
\be\label{eq:smalldelta}
I_\text{osc.} =  \int_0^\infty d\delta \int_{-\delta}^\delta d \nu \frac{1}{4 \delta^2}\left(\Big[\sin \Big( 2 \delta \, \tau(\l_1) \Big) - \sin \Big( 2 \delta \, \tau(\l_2)\Big) \Big]^2 - 1 \right) \,.
\ee
Here we defined the `time of flight' as in (\ref{eq:exp})
\be\label{eq:tau}
\tau(\l)  = - \frac{1}{\sqrt{2}} \int^{\l}_{\lambda_\star(\nu)} \frac{d\l'}{\sqrt{-V(\lambda') - \nu}} = 
- \log \frac{\l + \sqrt{\l^2 - 2 \mu}}{\sqrt{2 \mu}} \,.
\ee

The integrals in (\ref{eq:smalldelta}) can be done exactly. The answer is, again in terms of the time of flight variable,
\be\label{eq:n1n2}
I_\text{osc.} = \gamma_E + \log \epsilon +\log \frac{\tau(\l_2) - \tau(\l_1)}{\tau(\l_2)+ \tau(\l_1)}
+ \frac{1}{2} \log \Big(16  \tau(\l_1) \tau(\l_2)  \Big) \,.
\ee
Here $\epsilon$ is a cutoff on the lower (i.e. near $0$) end of the $\delta$ integral. This divergence will be cancelled by the non-oscillatory part of the integral to be considered shortly. The integral is convergent towards large $\delta$. Indeed, we see in (\ref{eq:smalldelta}) that the dominant contribution to the integral comes from $\delta \lesssim [\tau(\l_{1/2})]^{-1} \lesssim 1$ (we are assuming here that $\l_1$ and $\l_2$ are not too close to $\sqrt{2 \mu}$, as is indeed required by WKB), showing that it is consistent to have isolated the small $\delta$ region in the way we have done. An expression closely related to (\ref{eq:n1n2}), for the WKB limit of the density two point function, can be found in \cite{Moore:1991sf}.

The above computation in fact goes through for electrons in any potential $V(\lambda)$. The result is always (\ref{eq:n1n2}) in terms of the time of flight variable for the potential. Furthermore, the contribution of these modes close to the Fermi surface to the higher cumulants $V^{(m)}$ in (\ref{eq:cumu}) is easily seen to be only a function of the time of flight variable (\ref{eq:tau}).

The non-oscillating contribution to the integral (\ref{eq:quartic}) is also transparently discussed in terms of a general potential $V(\lambda)$. Once the $\lambda$ integral has been performed using (\ref{eq:trick}) above, the non-oscillating part of the integrand for the $\nu_1$ and $\nu_2$ integrals is $I(\l_1) + I(\l_2)$, where
\be\label{eq:II}
I(\l) \equiv \frac{1}{4 \left(\nu_1 - \nu_2 \right)^2} \left(\sqrt{\frac{-V(\l) - \nu_1}{-V(\l) - \nu_2}} + \sqrt{\frac{-V(\l) - \nu_2}{-V(\l) - \nu_1}}\right) \,.
\ee
We only kept terms here that are leading order in the WKB expansion. The remaining terms vanish as $\mu \to \infty$. By again transforming to $\nu$ and $\delta$ coordinates as in (\ref{eq:nudelta}), the integrals can be performed analytically. Performing the $\nu$ integral leaves
\be\label{eq:din}
\int_0^\infty \frac{d \delta}{4 \delta^2} \left( \sqrt{-V(\l_1)} \left(\sqrt{- V(\l_1) + 2 \d} - \sqrt{- V(\l_1) - 2 \d} \right) + ( \l_1 \leftrightarrow \l_2 ) \right) \,.
\ee
Rather than doing this integral explicitly, we need to think at this point about the limits of integration. Firstly, note that there is a divergence as $\d \to 0$. If we put in a lower cutoff at $\epsilon$, we see that the divergent contribution is
\be
 - \log \epsilon \,,
\ee
which exactly cancels the divergence we found in (\ref{eq:n1n2}). Therefore, the full answer has no divergence in the contribution from near the Fermi surface. This is a crucial part of our result which ensures that the entanglement entropy is manifestly finite.

The integrand in (\ref{eq:din}) becomes complex for $2 \delta > - V(\l_{1/2})$. This is because the integral goes outside the regime in which the oscillating WKB wavefunction (\ref{eq:WKB}) is valid. Outside the oscillating region, the wavefunction is exponentially small. The trick we used in (\ref{eq:trick}) remains valid in all regions. Therefore, all we have to do is set the wavefunction to zero whenever one or both of the limits of the $\lambda$ integral in (\ref{eq:trick}) is in the classically forbidden region. 
Taking this into account, the total expression becomes
\be
I_\text{non-osc.} = \int_{-\infty}^\mu d \nu_2 \left( \int_{\mu}^{\l_1^2/2} d \nu_1 I(\l_1) + \int_{\mu}^{\l_2^2/2} d \nu_1I(\l_2) \right) \,,
\ee
with $I(\l)$ given in (\ref{eq:II}). These integrals are easily done and we obtain (with $V(\lambda) = -\lambda^2/2$)
\be
I_\text{non-osc.} = \frac{1}{2} \log \left[ \left( \lambda_1^2 - 2\mu \right)\left( \lambda_2^2 - 2\mu \right) \right] - \log \epsilon \,.
\ee

Putting the above together (i.e. $3 S_A = I_\text{osc.}+I_\text{non-osc.}$), and using the inverse relation to (\ref{eq:tau}), i.e.
\be
\lambda = \sqrt{2 \mu} \cosh \tau(\lambda) \,.
\ee
we finally obtain the entanglement entropy
\bea
3 S_A & = & \log \frac{\tau(\l_2) - \tau(\l_1)}{\tau(\l_2) + \tau(\l_1)}
+ \frac{1}{2} \log \Big(\tau(\l_1) \tau(\l_2)  \Big) + \gamma_E + \log 4 \nonumber \\
&  & \qquad + \, \log \Big( 2 \mu \sinh\tau(\lambda_1) \, \sinh\tau(\lambda_2)  \Big) \,. \label{eq:SS}
\eea
This is our answer. From the point of view of the collective bosonized excitations of the Fermi surface,
it is natural to introduce the `string coupling' \cite{Jevicki:1993qn}
\be\label{eq:gtilde}
\frac{1}{\widetilde g_\text{s}(x)} \equiv 2 \mu \sinh^2 \tau(\lambda) \,.
\ee
Under the identification (\ref{eq:exp}), in which $\tau(\l) = x$, this expression does not map directly onto the target space string coupling (\ref{eq:gs}). However, in the limit in which the expression (\ref{eq:gs}) for the coupling can be trusted, $x \to - \infty$, then indeed we have
\be
\widetilde g_\text{s}(x) = \frac{g_\text{s}(x)}{2 \mu} \,, \qquad (x \to -\infty) \,.
\ee
Using (\ref{eq:gtilde}) we can rewrite (\ref{eq:SS}) as
\be\label{eq:final}
\fbox{
$
\displaystyle
S_A = \frac{1}{3} \log \frac{x_2 - x_1}{\sqrt{\widetilde g_\text{s}(x_1) \, \widetilde g_\text{s}(x_2)}} + \frac{1}{6} \log \frac{16 \, x_1 x_2}{(x_1 + x_2)^2} + \frac{\gamma_E}{3} + \cdots \,.
$
}
\ee
The dots denote the contribution from the higher cumulants in (\ref{eq:cumu}), that we discuss shortly.
In this formula we used the identification (\ref{eq:exp}) to set $\tau(\lambda_1) = x_2$ and $\tau(\lambda_2) = x_1$ (we flipped the numbering $1 \leftrightarrow 2$ to preserve the ordering under the change of sign as $\lambda \leftrightarrow x$).
This is our final expression for the entanglement entropy of the region $[x_1,x_2]$ in the target space theory.
It is manifestly finite. We noted in (\ref{eq:13}) and (\ref{eq:result}) that the result reproduces the expected
entanglement due to the tachyon field in the emergent spacetime, in the regime in which the comparison can
be made reliably.

We have already mentioned the results in \cite{calab1, calab2}: that the leading logarithmically singular term in the entanglement entropy at large particle number is entirely captured by the second cumulant $V_A^{(2)}$. We argued that for a region $[\lambda_1,\lambda_2]$ that is very small compared to the scale of variation of the potential, this result should hold for fermions in any potential. To leading order this singular term would go like $\log (\lambda_2 - \lambda_1)$. We have seen that this term arises due to modes close to the Fermi surface. We furthermore noted that the contribution of these modes to all the $V_A^{(m)}$ cumulants are functions of the time of flight variables $\tau(\lambda)$. Taken together, these observations suggest that the whole $\log [\tau(\l_2) - \tau(\l_1)]$ term found in (\ref{eq:SS}) is reliable. This is a stronger statement. We have checked this stronger statement by verifying the absence of a $\log [\tau(\l_2) - \tau(\l_1)]$ term in the first higher cumulant correction, $V_A^{(4)}$. Therefore, we expect that the first term in the final result (\ref{eq:final}) correctly captures the leading non-analyticity in the entanglement entropy as $x_1 \to x_2$ at any $x$. We do not have an analogous argument for the remaining terms in (\ref{eq:final}). Numerical investigation of the $V_A^{(4)}$ correction suggests that these terms do receive corrections, and hence are not reliable, although it is possible that they describe additional structure in the entanglement due to the nonlinear physics of the tachyon field coupled to the non-uniform background tachyon condensate.

The cutoff in the short distance target space entanglement, as evidenced in the
comparison of (\ref{eq:13}) and (\ref{eq:result}), arises technically from the fact that the inverse string coupling (\ref{eq:gtilde})
is essentially the depth of the Fermi sea at the point $\lambda$. This finite depth provides a cutoff on the number of modes
that are entangled, as was emphasized in \cite{das1,das2}. Therefore, the cutoff in the first term in (\ref{eq:final}) should also be robust, to leading order in the WKB limit, against corrections from the higher cumulants \cite{calab1, calab2}. We have explicitly checked that there is no logarithmically large contribution to the non-oscillating part of $V_A^{(4)}$.

\section{Discussion}

We have shown that the emergent semiclassical locality in two dimensional string theory can indeed by seen by partitioning the underlying quantum mechanical Hilbert space in the expected way. Namely, the degrees of freedom in some interval of the emergent target space are described by the eigenvalues of a matrix quantum mechanics that take values in some corresponding range. By explicitly
calculating the entanglement entropy, we identified the large amount of short distance entanglement expected due to emergent locality. We found that this accumulation of entanglement is cut off, from the spacetime point of view, by a nonperturbative `graininess' at a scale set by the (spatially varying) string coupling constant.

The emergent locality that we have diagnosed through entanglement goes all the way to the spacetime UV cutoff scale.
Recent works in the AdS/CFT correspondence have noted that an emergent coarse-grained locality, at the larger AdS scale, can
be represented by the entanglement structure of a tensor network \cite{Swingle:2009bg, Qi:2013caa, Almheiri:2014lwa, Pastawski:2015qua}. This `skeleton' of the emergent spacetime needs to be fleshed out with a large $N$ number of degrees of freedom, that can then provide locality of the sort we have described: down to a microscopic scale. The single matrix quantum mechanics we have studied is not powerful enough to produce spacetime physics with a hierarchy of scales. We hope that a similar perspective to the one we have taken can be applied also to more complicated theories of matrix quantum mechanics.

Our investigation was oriented towards using entanglement as a diagnostic of emergent locality.
We argued that the entanglement entropy we computed should be understood as the entanglement
of the tachyon field in the emergent space. In physically richer theories of quantum gravity,
the entanglement of fields in the emergent spacetime will come hand in hand with the
entanglement of microscopic `stringy' degrees of freedom that act as the `architecture of spacetime'
\cite{Susskind:1994sm, Fiola:1994ir, Bianchi:2012ev, Faulkner:2013ana}. Thus, the understanding of
emergent locality in fully fledged quantum gravity theories, using the perspective advocated here,
will likely be closely tied to a deeper understanding of the Bekenstein-Hawking entropy of black holes
\cite{Bekenstein:1973ur, Hawking:1974sw} as well as the Ryu-Takayanagi formula \cite{Ryu:2006bv}.
The singlet sector matrix quantum mechanics we have studied is not expected 
to capture black hole thermodynamics nor, likely, Ryu-Takayanagi type spacetime entanglement.
See, for instance, the discussion in \cite{Karczmarek:2004bw} and references therein.

Still within two dimensional string theory, part of the power of the free fermion description is that the Schr\"odinger equation for non-interacting fermions in an inverted harmonic oscillator can be solved exactly in terms of confluent hypergeometric functions, without needing to take the $\mu \to \infty$ WKB limit. Study of the entanglement entropy in this more general case offers to shed light on the spacetime string theory away from the limit of weak interactions. While locality is likely to become blurry in this case, the linear dilaton background gives a diffeomorphism-invariant `labeling' of points in space.

\section*{Acknowledements}

We are happy to acknowledge useful discussions with Matt Headrick, Juan Maldacena and Lenny Susskind.
This work was partially supported by a DOE Early Career Award, by the Templeton Foundation and an NSF Graduate Research Fellowship.

\end{document}